\documentclass[reprint,superscriptaddress,amsmath,amssymb,aps,longbibliography]{revtex4-1}
\usepackage{graphicx}
\begin{document}

\title{
Jamming, Fragility and Pinning Phenomena in Superconducting Vortex Systems
} 
\author{
C. Reichhardt and C. J. O. Reichhardt 
} 
\affiliation{
Theoretical Division and Center for Nonlinear Studies,
Los Alamos National Laboratory, Los Alamos, New Mexico 87545, USA\\ 
} 

\date{\today}

\maketitle

{\bf
We examine driven superconducting vortices interacting with quenched disorder where we measure the memory effects under a sequence of drive pulses applied perpendicular to each other.  As a function of disorder strength, we find four types of behavior: an elastic response, a fragile phase, a jammed phase, and a pinning dominated regime.  These phases can be distinguished by the presence or absence of different types of memory effects.  The fragile and jammed states exhibit memory, while the elastic and pinning dominated regimes do not.  In the fragile regime, the system organizes into a pinned state during the first drive pulse, flows during the second drive pulse which is applied in the perpendicular direction, and then returns to a pinned state during the third drive pulse which is applied in the same direction as the first pulse.  This behavior is the hallmark of fragility that has been proposed for jamming in particulate matter by Cates {\it et al.} [Phys. Rev. Lett. 81, 1841 (1998)].  For stronger disorder, we observe a robust jamming state in which a memory effect is still present but the system reaches a pinned or reduced flow state during the perpendicular drive pulse, similar to what appears in granular systems under shear jamming.  We show that the different states produce signatures in the spatial vortex configurations, with memory effects arising from the coexistence of an elastic component and a pinned component in the vortex assembly.  The sequential  perpendicular driving protocol we propose for distinguishing fragile, jammed, and pinned phases should be general to the broader class of driven interacting particles in the presence of quenched disorder.
}

The jamming phenomenon refers to the cessation of flow in an assembly
of interacting particles due to the formation of particle configurations
which are stable against further motion even in the presence of a drive
or load
\cite{Liu98,Cates98,OHern03,Liu10}.
When the particle-particle interactions are short ranged, such as for
grains or bubbles,
there can be a well-defined specific density at which 
jamming occurs called point $J$
\cite{Liu98,OHern03,Liu10,Drocco05,Olsson07,Reichhardt14}. 
When friction or additional constraints are introduced,
such systems can jam at much lower densities under a drive
and can exhibit two different regimes, the first of which
is called a fragile phase where jamming occurs in the direction of
shear but not in other directions
\cite{Cates98,Bi11,Seto19}.
In the second regime,
driving can induce the formation of
jammed phases such as shear jammed states \cite{Bi11,Sarkar16,Otsuki18}.

A phenomenon related to jamming is
clogging,
where the particle flow ceases due to a
combination of particle-particle interactions and
physical constraints such as a wall or posts, as
observed in
the flow of particles through a hopper \cite{To01,Thomas13},
bottlenecks \cite{Zuriguel15}, channels \cite{Wyss06,Barre15}, 
and obstacle arrays
\cite{Barre15,Chevoir07,Reichhardt12,Graves16,Nguyen17,Peter18,Reichhardt18a,Stoop18}. 
A unifying 
theme in jamming and clogging  
is that particles must reach
specific configurations in which the interactions create a network
that can stop the flow.
One consequence
of this is that some jammed states
can exhibit fragility, as introduced by
Cates {\it et al.} \cite{Cates98},
where load-bearing force chains block the flow for the driving direction
but not in other directions.
The ideas of jamming and fragility have generally been applied
to systems with short range or contact-only forces such as
granular matter, emulsions and colloids; 
however, some of these ideas could
be extended to system with longer range interactions.      

Another type of system that exhibits
transitions from flowing to immobile states is
assemblies of particles interacting
with quenched 
disorder or pinning \cite{Fisher98,Reichhardt17}. Examples
of such systems include vortices in type-II superconductors
\cite{Bhattacharya93,Blatter94}, 
electron crystals \cite{Williams91,Reichhardt01},
skyrmions \cite{Iwasaki13,Reichhardt15}, emulusions \cite{LeBlay20}, 
colloids \cite{Pertsinidis08,Tierno12a},
and active matter \cite{Chardac20}. In these systems, 
the motion can be halted not only due to direct interactions between
the particles and the pinning sites, but also by the blocking of
unpinned particles through interactions with directly pinned particles.
As a result,
interactions 
play a significant role in the pinned state
just as they do in the formation of jammed states.
There are different types of depinning phenomena
including elastic depinning, where the particles
form a rigid solid with no exchanges or plastic events, and
plastic depinning, where the particles
can exchange places or tear past one another at
and above depinning \cite{Fisher98,Reichhardt17}.
The transition from plastic
flow to a pinned state should have
the most similarities to a jamming transition from a mixing fluid state
to a solid non-flowing state.  

Despite the common features exhibited by both jamming and pinning,
a connection between these two phenomena has not yet been established.
One of the most studied systems that exhibits depinning is vortices
in type-II superconductors,
which can exhibit both plastic and elastic depinning transitions
\cite{Fisher98,Reichhardt17,Bhattacharya93,Blatter94}.
Driven vortices also exhibit many
memory effects, in which 
the depinning threshold is strongly affected by the manner in which
the pinned vortex configuration was created
\cite{Henderson96,Xiao99,Banerjee99,Xiao01,Du07,Pasquini08,Okuma12,Shaw12}. 
More recently it was shown that
vortices can exhibit reversible to irreversible
transitions or organize to specific dynamical states under 
periodic drives \cite{Mangan08,Okuma11,Dobroka17,Bermudez20},
similar to the behavior found
in colloidal systems and amorphous jammed systems
\cite{Corte08,Tjhung15,Regev13,Fiocco14,Keim14,Priezjev16},
suggesting that vortices can have properties
resembling those in other systems that 
exhibit jamming or fragility.

\begin{figure}
  \begin{center}
\includegraphics[width=\columnwidth]{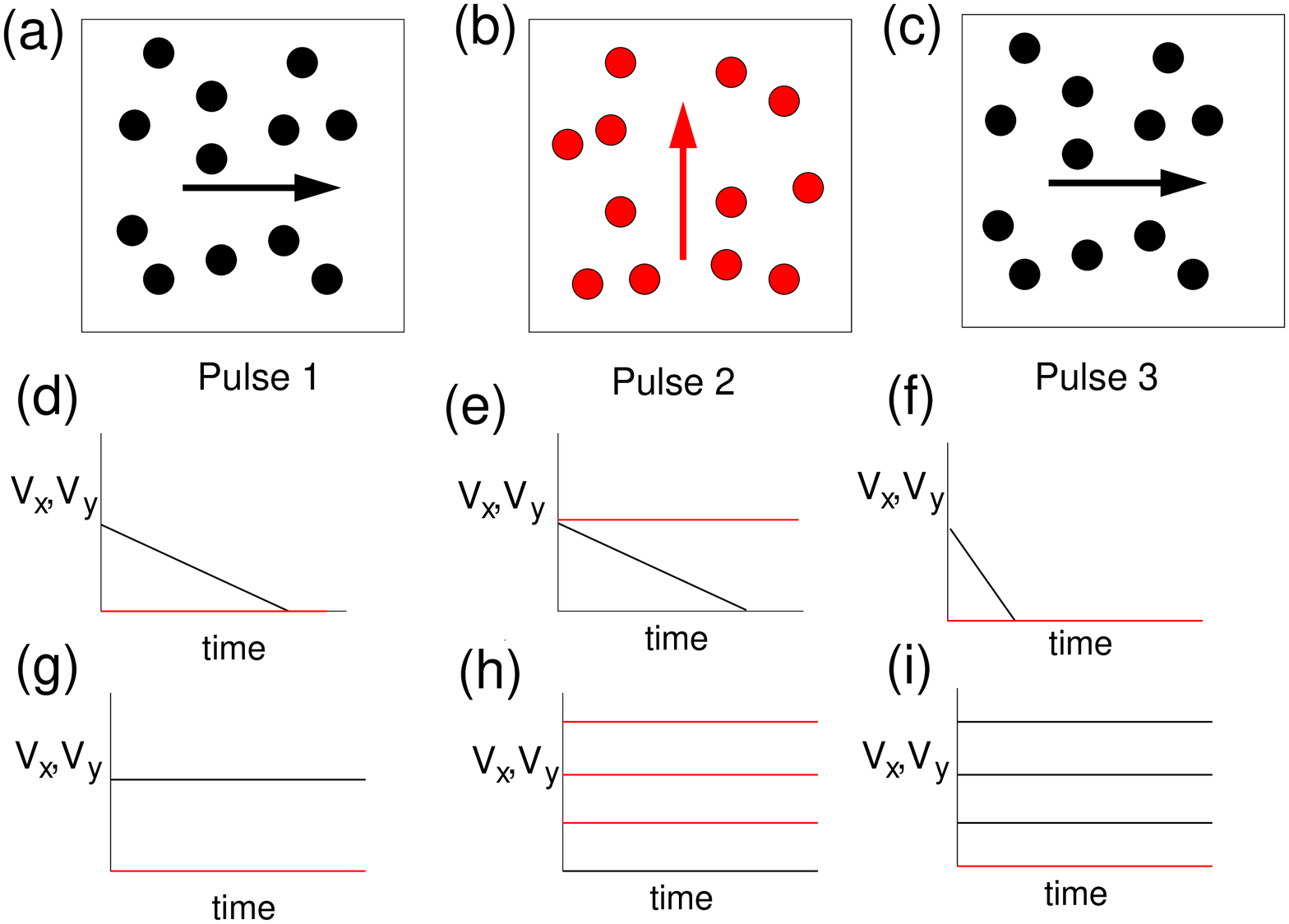}
\caption{ {\bf 
Schematic illustration of the driving protocol}.
{\bf a}, Pulse 1 ($p_1$) is applied along the $x$ direction,
and if the system organizes into a pinned state
the velocity response time series
appears similar to that plotted in panel {\bf d}, where
the $x$-direction velocity $V_x$ (black curve) decays to zero and the
$y$-direction velocity $V_y$ (red curve) is always zero.
{\bf b}, Pulse 2 ($p_2$) is then applied along the $y$ direction, giving
a distinctive velocity response such as that plotted in panel {\bf e}.
{\bf c}, Finally pulse 3 ($p_3$) is applied along the $x$ direction,
producing a velocity response of the type shown in panel {\bf f}.
If, during pulse $p_1$, the steady state $V_x$ is finite, as shown
in panel {\bf g}, then possible velocity responses during $p_2$ and $p_3$
are plotted in panels {\bf h} and {\bf i}, respectively.
If the system has no memory, the response in the driving direction
will be the same during all three pulses.
}
\label{fig:1}
\end{center}
\end{figure}

An open question is how to distinguish between jamming and pinning-dominated
behaviors in superconducting vortices or in general systems
with quenched disorder.
Here we
propose a method to use a pulsed drive protocol to
characterize these different phases 
for vortices interacting with random disorder.
We first apply a drive pulse to the system in one direction for a
fixed period of time,
than shut off that drive and
apply a second drive pulse in the perpendicular direction for the
same duration of time.  We then shut off the second pulse and
apply a third pulse in the original 
direction to test whether the system retained any memory
of the first driving pulse.
In Fig.~\ref{fig:1}a,~\ref{fig:1}b, and~\ref{fig:1}c we illustrate a schematic of the
driving protocol.
During pulse 1, termed $p_1$, the drive is applied in the $x$
direction, and Fig.~\ref{fig:1}d shows
a time series of the possible velocity
responses $V_x$ and $V_y$ in the $x$ and $y$ directions for the case
where the system organizes into a pinned state.
Here $V_y$ is always zero and $V_x$ decays to zero over some time
interval.
The second driving pulse $p_2$ is applied along the $y$ direction,
and the response along the $y$ direction can either be
immediate formation of a pinned state, initial flow followed by
organization into a pinned state, or a steady flow phase with
finite $V_y$, as shown
schematically in Fig.~\ref{fig:1}e.
If $p_1$ produces a pinned state but $p_2$ causes steady flow, this is an
indicator of fragility.
During pulse $p_3$, applied along the $x$ direction,
the same three responses of immediate pinning, organization to a pinned
state (as plotted schematically in Fig.~\ref{fig:1}f), or steady flow
are possible.  
If the system reaches steady state flow during $p_1$, as in
the schematic
in Fig.~\ref{fig:1}g,
then during $p_2$ several velocity responses are possible, as shown in
Fig.~\ref{fig:1}h.
These include greater flow during $p_2$ compared to $p_1$,
indicating a fragile state,
reduced flow during $p_2$ or formation of a pinned state, indicating jamming,
or the same amount of flow in $p_1$ and $p_2$,
indicating that the system has no memory.
Finally Fig.~\ref{fig:1}i shows that for this case, the response
during $p_3$ can either be pinned, reduced flow compared to $p_1$, greater
flow than in $p_1$, or flow identical to that in $p_1$.

We find that when the number of vortices is greater than the number
of pinning sites, two different types of jamming behaviors appear.
For weak disorder,
the system exhibits fragility.  Here, during $p_1$ the vortices organize into
a pinned state in which vortices directly trapped by the pinning sites
block the flow of the interstitial vortices that are in the regions between
the pinning sites.
This configuration is unstable to a change in the driving direction,
so during $p_2$ a flowing state forms
since a portion of the vortices that became directly pinned during $p_1$
remain trapped and
cannot rearrange to create a new network that could block the flow in
the $y$ direction.  
In this fragile case, during $p_3$ when the
drive is reapplied in the original $x$ direction,
the system retains a memory of the initially pinned state
and the vortices are able to recover
the same pinned state formed during $p_1$. 
For stronger disorder, we observe a more robust jamming behavior
similar to that found in shear jammed states.
Here, even if $p_1$ resulted in steady state flow,
the application of $p_2$
produces a lower flow mobility,
and under repeated perpendicular switching of the driving direction,
the system eventually reaches a fully jammed state.
In the pinning-dominated regime,
if the system reaches a pinned state during $p_1$, then it
remains pinned during $p_2$ and during all subsequent drive pulses.
If the drive is increased above the depinning threshold,
flow occurs during all drive pulses, but if the velocity response
is anisotropic,
with lower flow for $p_1$ and higher flow for
$p_2$, the behavior is fragile.
If the flow is instead smaller during $p_2$ than during $p_1$,
this is an indication that jamming behavior would occur at
lower drives.
We also find that for weak pinning or low pinning density,
the memory disappears and the response is more elastic,
with rearrangements occurring among the directly pinned vortices.

We map the regions in which fragile and pinned phases
occur as a function of drive strength and pinning density.
We also discuss how these
results could be general to other systems
that contain quenched disorder or that exhibit nonequilibrium memory effects.   

{\it Simulation and System---} 
We consider
a two-dimensional $L \times L$ system of rigid vortices
with periodic boundary conditions in the $x$ and $y$-directions.
The system contains
$N_{v}$ vortices and $N_{p}$ pinning sites, with
vortex density $n_v=N_v/L^2$, pinning density $n_p=N_p/L^2$, and
a vortex to pin ratio of $R=n_v/n_p$.
The vortex dynamics is obtained using
the following overdamped equation of motion: 
\begin{equation} 
\alpha_d {\bf v}_{i}  =
{\bf F}^{vv}_{i} + {\bf F}^{p}_{i} + {\bf F}^{D} .
\end{equation}
The vortex velocity is
${\bf v}_{i} = {d {\bf r}_{i}}/{dt}$
and $\alpha_d$ is the damping constant
which we take equal to 1.
The vortex-vortex interactions are given by
${\bf F}^{vv}_{i} = \sum^{N}_{j=1}F_{0}K_{1}(r_{ij}/\lambda){\hat {\bf r}_{ij}}$,
where $F_{0} = \phi^2_{0}/2\pi\mu_{0}\lambda^3$, 
$\phi_{0}$ is the elementary flux quantum, $\lambda$ is the London
penetration depth,
and $\mu_{0}$ is the permittivity.
The distance between vortex $i$ and vortex $j$ is
$r_{ij} = |{\bf r}_{i} - {\bf r}_{j}|$
and $K_{1}$ is the modified Bessel function, which falls off exponentially
for large $r_{ij}$.
For $r_{ij}/\lambda > 6.0$ 
the interaction between vortices is weak enough that it can be
cut off for computational 
efficiency, and larger cutoffs
produce negligible differences in the results
for the range of vortex densities we consider. 
In the absence of pinning,
the vortices form a triangular lattice with lattice constant $a$.  
The $N_{p}$ pinning sites are placed in random
but non-overlapping positions.
Each pinning
site is modeled as a parabolic trap of
range $r_{p}$ that can exert a maximum pinning
force of $F_{p}$ on a vortex.
The vortex-pin interaction
has the form
${\bf F}_i^{p} = \sum_j^{N_p}(F_{p}/r_{p})(r_{i} -r_{j}^{(p)})\Theta(r_{p} -|{\bf r}_{i} 
- {\bf r}_{j}^{p}|)$.
Our unit of length is $\lambda$ and we set
$L = 36$, $F_p=1.0$,
and $r_{p} = 0.25$.
We focus on the case of a vortex density of
$n_{v}= 1.67$ with $N_v=2160$
vortices and vary the pin density from
$n_p=0$ to $n_p=2.65.$  

The initial vortex configurations are obtained by starting
the system at a high temperature and gradually cooling
to $T = 0.0$.
After annealing, we apply a series of constant drives to
the vortices,
${\bf F}_D=F_D{\hat{\alpha}}$,
where $\hat{\alpha}={\bf \hat{x}}$ during
drive pulses $p_1$ and $p_3$ and $\hat{\alpha}={\bf \hat{y}}$ during
drive pulse $p_2$.
We measure the average velocity
per vortex in the $x$ and $y$ directions,
$\langle V_{x}\rangle = N_v^{-1}\sum^{N_v}_{i=1} {\bf v}_{i}\cdot {\bf \hat{x}}$
and $\langle V_{y}\rangle = N_v^{-1}\sum^{N_v}_{i=1} {\bf v}_{i}\cdot {\bf \hat{y}}$. 
Drive pulse $p_1$ is applied along the $x$-direction for a duration of 
$\tau = 2.2\times 10^6$
simulation time steps.
During drive pulse $p_2$, we shut off the $x$ direction driving
and apply a drive of the same
magnitude $F_D$ along the $y$ direction
for the
same duration $\tau$.
If the system has no memory of the initial drive, 
then the average velocity {\it in the driving
direction} will be equal during both pulses,
$|\langle V_x\rangle^{1}| = |\langle V_y\rangle^{2}|$. 
After the first two pulses, we can 
continue the pulse sequence
by
applying a third pulse $p_3$ in the same direction as $p_1$ to test for
additional memory effects.

{\it Results--}
\begin{figure}
  \begin{center}
\includegraphics[width=\columnwidth]{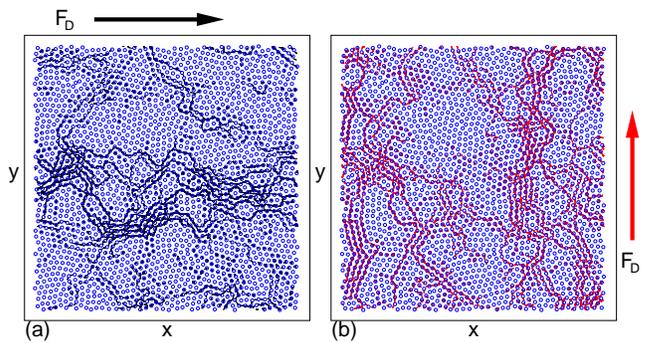}
\caption{
  {\bf Fragile phase flow.}
  Vortex positions (circles) and trajectories (lines)
in a system with $n_{p} = 1.0$.
The pinning site locations are not shown.
We apply a drive pulse $p_1$ of magnitude $F_D=0.045$ along the $x$ direction
(black arrow) for
a duration of $\tau$ simulation time steps,
followed by a drive pulse $p_2$ of the same magnitude and duration
along the $y$ direction (red arrow).
{\bf a}, Motion during $p_1$, where there is a
combination of pinned and flowing vortices.
{\bf b}, Motion during $p_2$, where the amount of flow is increased and the flow
direction has rotated to follow the drive.
}
\label{fig:2}
\end{center}
\end{figure}

In Fig.~\ref{fig:2} we plot the vortex locations and trajectories
over a fixed time period
for a system with
$F_{p}= 1.0$ and $n_{p} = 1.0$ under driving of magnitude $F_D=0.045$.
During pulse $p_1$ in Fig.~\ref{fig:2}a,
there is a combination of pinned and flowing vortices, and the motion
is mostly along the $x$ direction, parallel to the drive.
During pulse $p_2$, which is applied
along the $y$-direction, Fig.~\ref{fig:2}b illustrates
that there is now a greater amount of motion in the system.
This indicates that when the drive is applied initially
in the $x$-direction,
the flow organizes to a partially jammed state in which a portion of the
flow is blocked in the $x$ direction but not in the $y$ direction.
During pulse $p_3$, the flow is identical to that shown in Fig.~\ref{fig:2}a.
If $F_{D}$ is reduced,
we find
states in which
the system organizes to a pinned configuration
during $p_1$ but is still able to flow
when the drive is switched to the $y$ direction during $p_2$.
For very low drives of $F_{D} < 0.03$,
the system becomes pinned for both driving directions.

\begin{figure}
  \begin{center}
\includegraphics[width=\columnwidth]{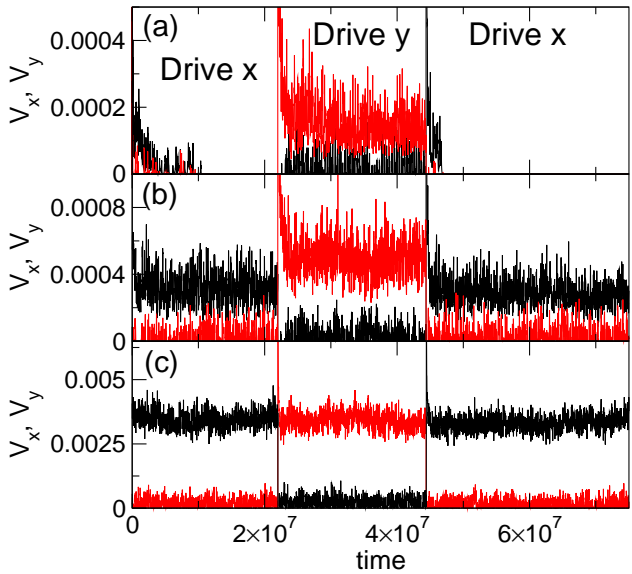}
\caption{
  {\bf Fragile phase velocity response.}
  The time series of the velocities $V_{x}$ (black) and $V_{y}$ (red)
for the system in Fig.~\ref{fig:2} with $n_{p} = 1.0$ for
pulse $p_1$ applied along the $x$ direction,
pulse $p_2$ applied along the $y$ direction,
and pulse $p_3$ applied along the $x$
direction within the fragile phase.
{\bf a}, At $F_{D} = 0.035$, the system organizes to a
jammed state during $p_1$, flows during $p_2$, and
returns rapidly to a pinned state during $p_3$.
{\bf b}, At $F_{D} = 0.045$, the system organizes
to a flowing state in $p_1$, exhibits a state with higher flow during $p_2$,
and returns to the first flow level during $p_3$.
{\bf c}, At $F_{D} = 0.08$, the memory effect is lost and the velocity in the
driving direction is the same during all drive pulses. }
\label{fig:3}
\end{center}
\end{figure}

To focus more specifically on the memory effects,
in Fig.~\ref{fig:3}a we plot the time series of $V_{x}$
and $V_{y}$ for the system in Fig.~\ref{fig:2} at
$F_{D} = 0.035$.
During pulse $p_1$,
$V_{y} = 0$ while $V_{x}$ initially has a finite value and gradually
drops to zero
as the system organizes into a pinned state.
When pulse
$p_2$ is applied, the $x$ direction motion remains absent with
$V_{x} = 0$,
but $V_{y}$ is finite and does not decay to zero.
This indicates that the system retains a memory of the initial
driving direction, and that 
the jammed state which forms for $x$ direction driving
is unstable against driving in a different direction,
which is a key signature of fragility \cite{Cates98}.
During pulse $p_3$,
when the drive is applied in the $x$-direction
again, the system returns to a pinned state
with $V_{x}=V_y=0$, indicating that a memory of the
initial driving in $x$ has persisted.
We note that if we continue to repeat alternating driving pulses in the
$x$ and $y$ directions,
there is always flow for $y$ direction driving
and no flow for $x$ direction driving.
The same system under a drive of magnitude
$F_D=0.045$ is illustrated in
Fig.~\ref{fig:3}b.
Here $V_{x}$ remains finite during pulse $p_1$
and $V_{y}$ is also finite during pulse $p_2$,
but the flow in the direction of drive is larger for the $y$ direction driving
than for the $x$ direction driving.
In this case, during pulse $p_1$, the system organizes to
a state in which only a portion of the flow
is jammed in the $x$ direction while
a smaller portion is jammed
in the $y$ direction.
This occurs because
many of the pinned vortices from $p_1$ remain pinned during $p_2$, as
shown in Fig.~\ref{fig:2}a,
so that even though there is flow for both driving directions,
there is still a
memory effect due to these pinned vortices.
During pulse $p_3$, $V_x$ returns to the same value it had
in $p_1$,
so the system remembers the flow state
that formed during the original driving pulse.
Figure~\ref{fig:3}c shows that in the same system at a higher drive
of $F_{D} = 0.08$,
$V_{x}$ in pulses $p_1$ and $p_3$ has the same value as $V_{y}$
in pulse $p_2$.
Here, many of the vortices are unpinned much of the time, causing the
destruction of the memory effect, which is replaced by a uniform
response that is insensitive to the driving direction.
When $F_{D} < 0.0325$, the system reaches a pinned state for both
directions of driving;
however,
the transient time required to reach the jammed
state during pulse $p_2$ in the $y$ direction is considerably
longer than that for the
initial pulse $p_1$ in the $x$ direction.

\begin{figure}
  \begin{center}
\includegraphics[width=\columnwidth]{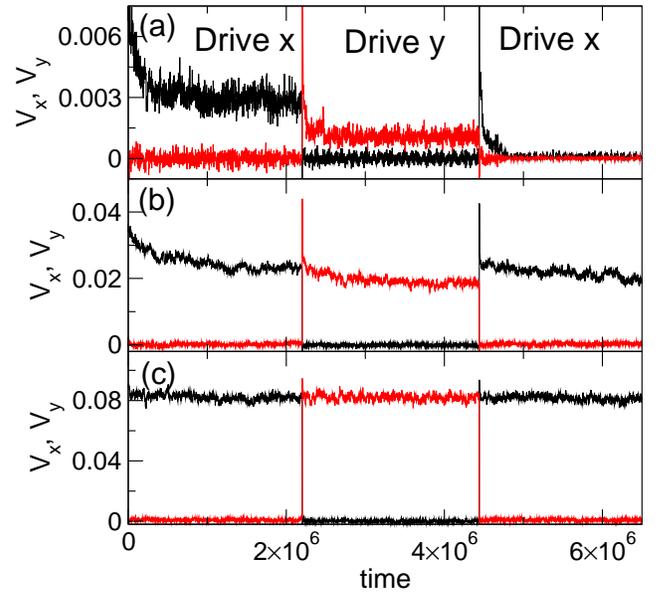}
  \caption{
  {\bf Jammed phase velocity response.}
The time series of the velocities
$V_{x}$ (black) and $V_{y}$ (red) for a system with $n_p=1.5$ where jamming
effects appear.  Driving pulse $p_1$ is applied along the $x$ direction, pulse
$p_2$ is applied along the $y$ direction, and pulse $p_3$ is applied along
the $x$ direction.
{\bf a}, At $F_{D} = 0.03$,
the system organizes to a flowing state during $p_1$, exhibits
reduced flow during $p_2$, and has almost no flow
during $p_3$.
{\bf b}, At $F_{D} = 0.475$, during $p_1$
the system organizes to a flowing state, followed by a state with
lower flow during $p_2$ and a return to the higher flow level
during $p_3$.
{\bf c}, At $F_{D} = 0.6$, $V_{x}$ for $x$ direction driving is equal
to $V_{y}$ for $y$ direction driving,
and the pulse memory effect is lost.
 }
  \label{fig:4}
  \end{center}
\end{figure}

The memory effect indicative of fragility,
where the response in the driving direction is larger during $p_2$
than during $p_1$,
occurs over the range
$0.45 < n_p < 1.25$ when
$n_v = 1.67$.
For $1.25 \leq n_{p} < 2.25$,
we observe a reversal of the memory effect in which
the system exhibits a more robust jamming effect
similar to that found in shear jammed systems.
In Fig.~\ref{fig:4}a we plot $V_{x}$ and $V_{y}$ as a function of time
for
a system with $n_{p} = 1.5$ at $F_{D} = 0.3$.
During pulse $p_1$,
$V_{x}$ decreases before reaching a finite steady state value.
When pulse $p_2$ is applied,
$V_{y}$ is lower than the steady state value of $V_x$ during pulse $p_1$,
indicating that rotation of the drive by 90$^\circ$ has allowed
the system to access a configuration with a higher amount of jamming.
During pulse $p_3$, $V_{x}$ drops to nearly zero, indicating a further
reorganization.
For lower $F_{D}$, the system can reach a pinned state during pulse $p_2$
and is completely pinned in pulse $p_3$.
In Fig.~\ref{fig:4}b,
for the same system at
$F_{D} = 0.475$, $V_y$ during pulse $p_2$ is lower than $V_x$ during
pulse $p_1$, but
during pulse $p_3$,
$V_x$ actually jumps up to a higher value
since almost all of the possible jammed configurations have
already been accessed by the system.
Figure~\ref{fig:3}c shows that at $F_{D} = 0.6$,
the response in the driving direction is the same during all drive
pulses
and the memory effect is lost.

\begin{figure}
  \begin{center}
\includegraphics[width=\columnwidth]{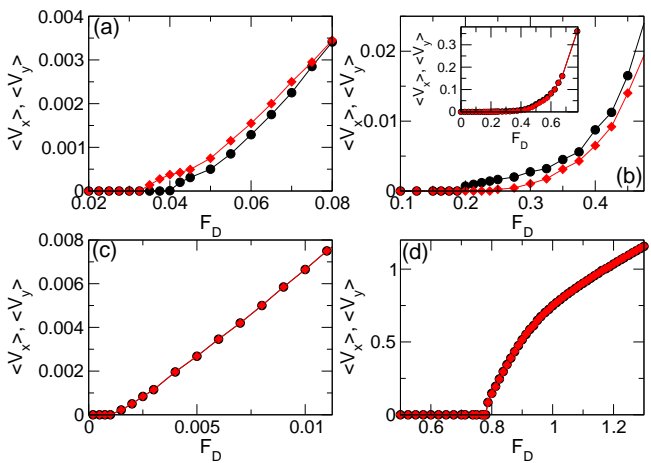}
\caption{{\bf Average velocity response under varied pulse magnitude.}
  $\langle V_{x}\rangle$ at the end of pulse $p_1$ (black circles)
and $\langle V_{y}\rangle$ at the end of pulse $p_2$ (red diamonds)
vs $F_{D}$.
{\bf a}, The fragile phase for the system in Figs.~\ref{fig:2} and \ref{fig:3}
with
$n_{p} = 1.0$, where $\langle V_{y}\rangle$ in $p_2$ is higher than
$\langle V_{x}\rangle$ in $p_1$.
{\bf b}, The jammed phase for the system in Fig.~\ref{fig:4} with $n_{p} = 1.5$
where $\langle V_{y}\rangle$ in $p_2$ is lower than $\langle V_{x}\rangle$
in $p_1$. The inset shows the same data out to higher values of $F_D$.
{\bf c}, The elastic phase for a system with $n_{p} = 0.1$,
where the response in the driving direction is the same
in both $p_1$ and $p_2$ and there is no memory effect.
{\bf d}, The pinning dominated phase
at $n_{p} = 2.625$, where there is no memory effect.
}
\label{fig:5}
\end{center}
\end{figure}

We can characterize the different phases by measuring the average velocity
$\langle V_x\rangle$ at the end of pulse $p_1$
and $\langle V_{y}\rangle$  at the end of pulse $p_2$ for varied $F_{D}$.
In Fig.~\ref{fig:5}a we plot
$\langle V_{x}\rangle$ and $\langle V_{y}\rangle$ versus
$F_{D}$ for the system in Figs.~\ref{fig:2} and \ref{fig:3}
with $n_p=1.0$ which is in the fragile phase.
Here, the system reaches a pinned state during both $p_1$ and $p_2$
when $F_{D} \leq 0.0325$,
while for
$0.0325 < F_{D} < 0.04$, the system jams during $p_1$
but flows during $p_2$.
For $0.04 \leq F_{D} < 0.08$,
both $\langle V_x\rangle$ and $\langle V_y\rangle$
are finite,
but the response during $p_2$ is larger than that during $p_1$.
The ratio $\langle V_{x}\rangle/\langle V_{y}\rangle$ for
pulses $p_1$ and $p_2$  gradually approaches
$\langle V_{x}\rangle/\langle V_{y}\rangle=1.0$
as $F_{D}$ increases, until
for $F_{D} \geq 0.08$, the response is the same for both driving
directions and the memory effect is lost.
In Fig.~\ref{fig:5}b, we plot $\langle V_x\rangle$ and $\langle V_y\rangle$
versus $F_D$ for the system in Fig.~\ref{fig:4} with
$n_{p} = 1.5$,
where jamming behavior appears.
For $0.1875 < F_{D} < 0.25$,
the system flows in the $x$ direction during pulse  $p_1$,
and is pinned in the $y$ direction during $p_2$.
When $0.25 \leq F_{D} < 0.6$, there is a finite response during
both $p_1$ and $p_2$, but the response is smaller in $p_2$.
For $F_{D} \geq 0.6$, the response becomes isotropic
as shown in the inset of Fig.~\ref{fig:5}b.
For samples with $n_{p} < 0.4$, the response becomes mostly elastic and
the memory effects disappear,
as shown in Fig.~\ref{fig:5}c where we plot $\langle V_{x}\rangle$
and $\langle V_{y}\rangle$ versus $F_{D}$ for a
system with $n_{p} = 0.1$.
The system is pinned for $F_{D} < 0.001$, but the response is isotropic
for higher $F_{D}$.
Samples with $n_{p} > 2.25$ are in the pinning-dominated regime,
where there is again
no memory effect, as shown in the plot of
$\langle V_{x}\rangle$ and $\langle V_{y}\rangle$ in Fig.~\ref{fig:5}d
for a system with $n_{p} = 2.625$.
In the pinning-dominated regime, almost every vortex can be pinned directly
at a pinning site, so the collective
effects associated with jamming or fragility are lost.

\begin{figure}
  \begin{center}
\includegraphics[width=\columnwidth]{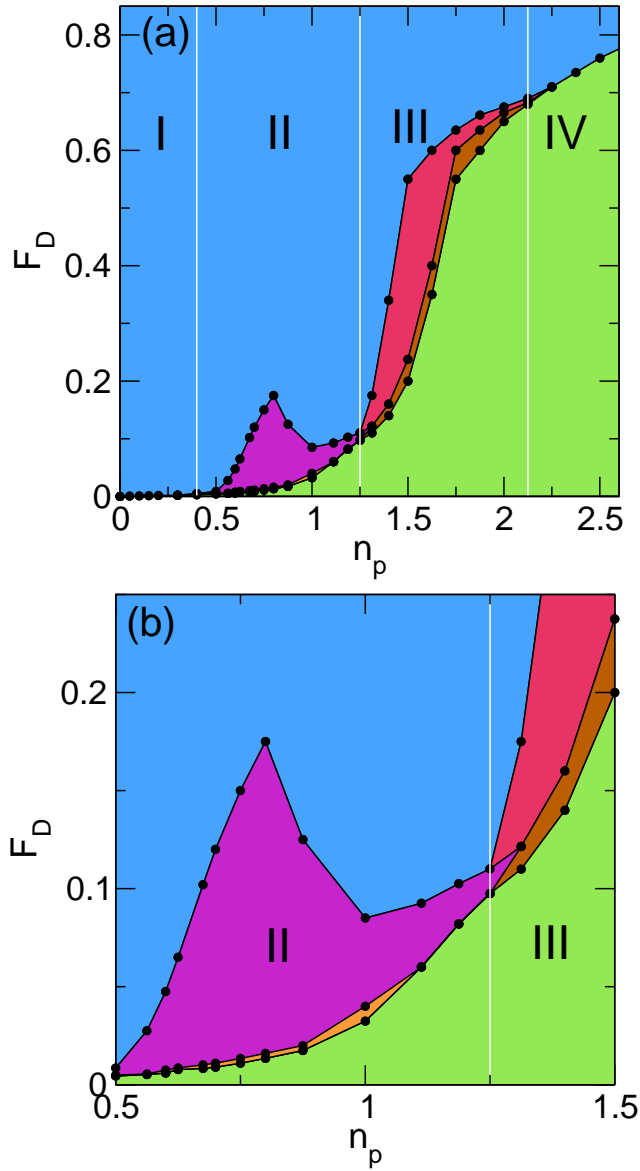}
\caption{
  {\bf Dynamic phase diagram as a function of $F_D$ vs $n_p$
obtained from a series of velocity versus pulse force measurements.}
{\bf a}, Green: Pinned phase where the system reaches a pinned
state during both $p_1$ and $p_2$.
Blue: Isotropic flow phase where the response in the driving
direction has the same value for both $p_1$ and $p_2$.
Orange: The fragile phase in which the system is pinned during $p_1$ and
flowing during $p_2$.
Purple: The fragile phase in which the system is flowing in $p_1$ and
has a larger magnitude of flow during $p_2$.
Brown: The jammed phase where the system flows in $p_1$ and is pinned
in $p_2$.
Red: The jammed phase in which there is flow in $p_1$ and reduced
flow in $p_2$.
The white lines indicate the divisions between the four phases that appear
as a function of pinning density below the isotropic flow regime.
Phase I: elastic regime with no memory.
Phase II: fragile jammed state with memory.
Phase III: robust jammed state with memory.
Phase IV: pinned state with no memory.
{\bf b}, A blow up of panel {\bf a} in the region of the fragile to jammed crossover.
}
\label{fig:6}
\end{center}
\end{figure}

By conducting a series of simulations for varied pulse magnitude $F_{D}$
and pinning density $n_{p}$, we can construct a dynamic phase diagram
highlighting the four different phases
as a function of $F_D$ versus $n_p$, as plotted in Fig.~\ref{fig:6}a.
In the pinned phase, the system reaches a pinned state during
both $p_1$ and $p_2$.
For the isotropic flow state, the velocity response in the driving
direction is the same during both $p_1$ and $p_2$.
The fragile response states include the regime in
which the system is pinned during $p_1$ and flowing during
$p_2$, as well as the parameter range in which the system is flowing during
$p_1$ and flowing faster during $p_2$.
The jammed phases consist of states in which
the system is flowing during $p_1$ but pinned during $p_2$ as well as
the regime in which there is flow during $p_1$ and reduced flow during $p_2$.
In Fig.~\ref{fig:6}b, we show a blowup of Fig.~\ref{fig:6}a
near the fragile to jamming crossover region to illustrate more clearly
the
different fragile and jammed phases.
The vertical white lines in Fig.~\ref{fig:6}a indicate the four
phases as a function of pinning density.
These are the elastic regime with no memory, marked as Phase I, which appears
for $n_{p} < 0.4$,
the fragile jammed state with memory, marked as Phase II,
for $0.4 \leq n_{p} < 1.25$,
the robust jamming state with memory, marked as Phase III, for
$1.25 \leq n_{p} < 2.125$,
and the pinned state with no memory, marked as Phase IV, for $n_{p} \geq 2.125$.
Near the boundary between fragile and jammed behavior,
multiple reversals in the velocity force curves can occur such that
the system may show a jamming behavior at lower drives but a fragile
behavior at higher drives.
At sufficiently high drives, in the blue region, the response becomes
isotropic.
We have constructed a similar phase diagram for a
vortex density of
$n_{v} = 0.75$ as a function of
$F_D$ versus $n_{p}$ and
find phase boundaries similar to those shown in Fig.~\ref{fig:6},
indicating
the robustness of the results.
In general, pinning dominated phases occur
when $n_{v}/n_{p} < 1.25$, where each vortex
can be trapped directly by a pinning site so that the vortex-vortex
interaction energies become unimportant.

\begin{figure}
  \begin{center}
\includegraphics[width=\columnwidth]{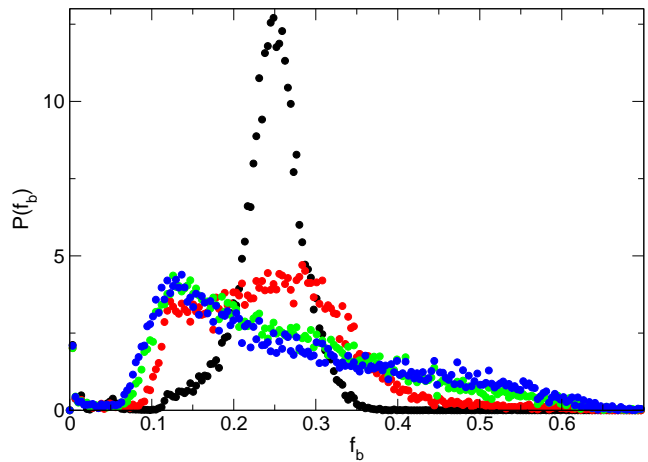}
\caption{
  {\bf Nearest neighbor forces in four phases.}
  Distribution $P(f_b)$ of nearest-neighbor vortex-vortex interaction
forces, histogrammed using five different realizations.
In the elastic phase I at $n_p = 0.1$ (black), there is a single narrow
peak in $P(f_b)$.  In the fragile phase II at $n_p=0.7$ (red), there are
two peaks on either side of a plateau,
and in both the jammed phase III at $n_p=1.5$ (green) and the
pinned phase IV at $n_p=2.5$ (blue) the distribution has a broad tail.
}
\label{fig:7}
\end{center}
\end{figure}

\begin{figure}
  \begin{center}
\includegraphics[width=\columnwidth]{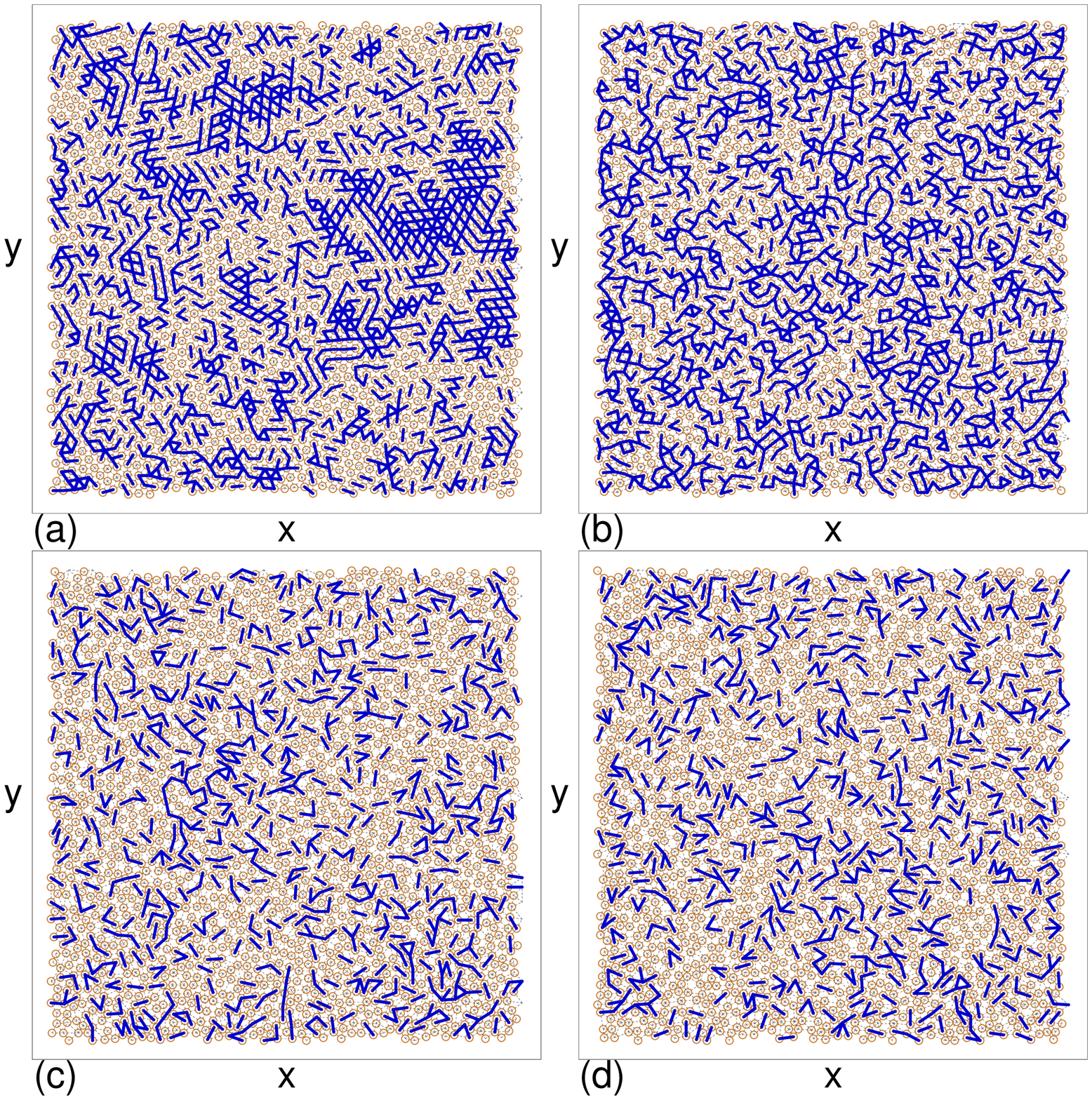}
\caption{
  {\bf Bond networks.}
  Images of the network of nearest-neighbor bonds between vortices
for example realizations of each system plotted in Fig.~\ref{fig:7},
showing the vortex positions (orange circles), bonds with a probability
of observation $P(f_b) \geq 0.9P_{\rm max}$
(heavy blue lines),
and the
remaining bonds (thin dashed black lines),
where $P_{\rm max}$ is the maximum
value of $P(f_b)$.
{\bf a}, The elastic phase I at $n_p=0.1$, where there are many
bonds with nearly the same strength since
the crystalline structure of the vortex lattice
prevents the vortices from adapting to the pinning landscape.
{\bf b}, The fragile phase II at $n_p=0.7$, where there is
a broad plateau in $P(f_b)$ that produces a highly adaptive bond network
that can retain a memory of the driving direction.
{\bf c}, The jammed phase III at $n_p=1.5$. 
{\bf d}, The pinned phase IV at $n_p=2.5$.
}
\label{fig:8}
\end{center}
\end{figure}

To gain insight into the onset of the different phases,
we measure the distribution of the forces
experienced by the vortices.
We first identify the true nearest neighbors of
each vortex using a Voronoi construction, and assign each nearest neighbor
pair to a bond.  Using the length of each bond, we compute the
interaction force $f_b$ between each pair of neighboring
vortices, and histogram this quantity using
five different realizations for Phases I through IV.
In Fig.~\ref{fig:7} we plot the distribution function $P(f_b)$
for the elastic phase I at
$n_{p} = 0.1$,
the fragile phase II at $n_{p} = 0.7$,
the jammed phase III at $n_{p} = 1.5$, and
the pinned phase IV at $n_{p} = 2.5$.
For the elastic phase we find a single relatively
narrow peak since most of the
vortices have six neighbors and the vortex-vortex spacing is fairly
uniform.
In Fig.~\ref{fig:8}a we show an image of the vortex-vortex nearest neighbor
bond network.  Highlighted in bold are those bonds whose frequency of
observation
$P(f_b)$ is within 10\% of $P_{\rm max}$,
$P(f_b) \geq 0.9 P_{\rm max}$, where $P_{\rm max}$ is
the maximum value of
$P(f_b)$. 
In the elastic regime, there are
large regions of sixfold ordering, producing large patches of bonds with
nearly the same value of $f_b$.  Since the vortex-vortex interactions
dominate in this regime, few rearrangements of the nearest neighbor
structure occur above the depinning transition during drive pulse $p_1$,
and the bond network structure remains unchanged during drive pulse $p_2$.  As a
result, the system has no memory of the driving direction.
In the fragile phase II, $P(f_b)$ has two peaks on either side of a broad
plateau, as shown in Fig.~\ref{fig:7} for a sample with $n_p=0.7$.
The peak at higher $f_b$ is associated with the remaining elastic component
of the vortex lattice, while the peak at lower $f_b$ arises due to increasing
plastic distortions of the lattice by the underlying pinning.
Due to the broad plateau, a wide range of values of $f_b$
meet the criterion $P(f_b)\geq 0.9 P_{\rm max}$,
producing a force
network of the type illustrated in Fig.~\ref{fig:8}b.
The large elastic patches that appeared in the elastic phase are replaced
by small clusters of bonds that have local orientation but
no global orientation.  These
disordered regions can undergo strong rearrangements in order to create
a percolating force network that
is able to stop the flow in the $x$ direction during drive pulse $p_1$.
A portion of these bonds break and rearrange
during drive pulse $p_2$,
but another portion of the bonds are preserved, so that when drive
pulse $p_3$ is applied, enough of the network that appeared in $p_1$ is still
present that the system is able to reform a similar network and
maintain a memory of its original driving direction.
In the jammed phase III, as shown in Fig.~\ref{fig:7} for $n_p=1.5$,
the elastic peak in $P(f_b)$ disappears and we
instead find a peak at lower $f_b$ along with a broad tail extending out
to high values of $f_b$.  Here the pinning interaction energy is beginning
to overwhelm the vortex-vortex interaction energy, producing some long
neighbor bonds with forces lower than those found in an elastic lattice
along with other short neighbor bonds with forces higher than those
that would be present in an elastic lattice.
Due to the highly skewed shape of $P(f_b)$, there are fewer bonds with
$P(f_b)\geq 0.9 P_{\rm max}$,
and these
bonds form a scattered and non-percolating structure, as illustrated
in Fig.~\ref{fig:8}c. Similar behavior of both
$P(f_b)$ and the bond network structure appears in the pinned phase IV,
as shown in Fig.~\ref{fig:7} and Fig.~\ref{fig:8}d for samples with
$n_p=2.5$.
In the jammed
phase, during successive drive pulses the system gradually approaches a
state resembling the pinned phase, where all motion has ceased and all
memory has disappeared.
The force distributions shown in Fig.~\ref{fig:7} have distinctive
signatures depending on whether the system is elastic, fragile,
or jammed/pinned.
As a result,
it may be possible
to determine whether 
elastic, fragile, jammed or pinned behavior is likely to occur
by examining the vortex configurations even before
a drive is applied.
This can be achieved experimentally using
a variety of imaging techniques \cite{Menghini02,Guillamon14,Ganguli15,Rumi19}.

{\it Discussion--}
Our results indicate that in experiments,
fragile or jamming behaviors would be most relevant for regimes
in which the number of vortices is greater than the number of
pinning sites or in samples where the pinning is strongly inhomogeneous.
Strong memory effects or metastable effects are generally
observed in superconductors that are either very clean or have only weak
pinning, which is consistent with our findings.
Jamming and fragility effects should be absent in or near the elastic
depinning regime,
where the vortex lattice is almost completely ordered,
as well as in systems with strong pinning.
Our results should be general to the broader class of
particle-like systems interacting with pinning,
including magnetic skyrmions, Wigner crystals, and
colloids with random pinning.
It is likely that similar
effects would arise in systems where the particle-particle
interactions are short ranged or have a hard sphere character, and
this will be explored in a future work.
The general protocol of applying a sequence of perpendicular drives
could also be used in systems without quenched disorder,
where it could provide a new method for studying
random organization processes of the type found
in periodic shearing of dilute colloidal systems.
For instance, this protocol could be applied to
dense amorphous soft solids to see if such systems also exhibit a
fragile to jamming crossover.
It is also possible that
different pulse protocols
could be used to determine whether the system retains different
types of memory,
similar to what has been observed in periodically sheared systems
\cite{Keim19}.

{\it Summary--}
We have proposed a protocol involving a sequence of
mutually perpendicular
driving forces
for determining whether a system with quenched disorder is
in an elastic, pinned, jammed, or fragile state.
In the fragile state,
the system is able to organize to a pinned state during the first drive pulse,
flows during the second drive pulse which is applied perpendicular to the
first pulse, but then returns
to the initial pinned configuration during the third pulse, where the drive
is applied in the original direction.
The fragile state is associated with a particle configuration that is
pinned only in the direction of drive but not in other directions.
In the jammed phase, the 
system can reach a steady state
flow during the first drive pulse,
but the application of the second pulse can cause
a reduction in the mobility,
while during the third pulse the system can cease flowing and reach
a jammed state
similar to the shear jamming behavior found in frictional granular materials.
For weak disorder, the memory effects are lost
since there are no longer any rearrangements of the particles with
respect to their nearest neighbors,
while for strong disorder or
the pinning dominated regime,
the memory is also lost since the response is dominated by
the pinning sites and not the vortex configurations.
Our results should be general to other classes of
particle-like systems interacting with quenched disorder
and even for other systems that exhibit random organization
or jammed phases.

%\bibliography{mybib}
%

\noindent{\bf Acknowledgements.}
This work was supported by the US Department of Energy through
the Los Alamos National Laboratory.  Los Alamos National Laboratory is
operated by Triad National Security, LLC, for the National Nuclear Security
Administration of the U. S. Department of Energy (Contract No. 892333218NCA000001).

\noindent{\bf Data availability} The data that support the plots within this paper and other findings of this study are available from the corresponding author upon request.

\noindent{\bf Author Contributions} All authors jointly conceived the project, performed the simulation, analyzed data and wrote the manuscript.

\noindent{\bf Author Information} Correspondence and requests for materials
should be addressed to C.~J.~O.~R. (email: cjrx@lanl.gov).

\noindent{\bf Competing interests} The authors declare no competing interests.

\end{document}